\documentclass{PoS}

\title{Gauge hierarchy and SUSY: Think again}

\ShortTitle{Gauge Hierarchy}

\author{\speaker{Jihn E. Kim}\thanks{This work is supported in part  by the National Research Foundation (NRF) grant  NRF-2018R1A2A3074631.}\\
         Department of Physics, Kyung Hee University, 26 Gyungheedaero, Dongdaemun-Gu, Seoul 02447, Republic of Korea  \\
         E-mail: \email{jihnekim@gmail.com}}

\abstract{We revisit the gauge hierarchy problem with the emphasis on the chiral property of the Standard Model. 
We  present a model realizing a gauge hierarchy.
Along this line, we also comment briefly on the ``very light'' axions and the upper bound on $\theta_{\rm QCD}$.}

\FullConference{Corfu Summer Institute 2019 "School and Workshops on Elementary Particle Physics and Gravity" (CORFU2019)\\
		31 August -- 25 September 2019\\
		Corfu, Greece}

\begin{document}
\section{Introduction}
 
The first mass scale defined in the lagrangian is the Planck scale $M_{P}\simeq 2.4\times 10^{18\,}$GeV which is known to be much larger than any mass scale used in several physics disciplines, 
\begin{eqnarray}
&&\textrm{Particle physics}:~246\,{\rm GeV}, \nonumber \\
&&\textrm{Intermediate scale physics}:~300\,{\rm MeV},\nonumber \\
&&\textrm{Nuclear physics}:~7\,{\rm MeV},\nonumber \\
&&\textrm{Atomic physics}:~1\,{\rm eV},\nonumber \\
&&\textrm{Condensed matter physics}:~10^{-3\,}{\rm eV}.\nonumber 
\end{eqnarray}
From particle physics scale, the smaller mass scales are obtained just by peeling the composite structures. So, understanding 246 GeV of particle physics is the key in understanding scales in all physics disciplines. But 246 GeV is about $10^{-16}$ times the Planck scale. To understand this ratio, we must solve the gauge hierarchy problem.

\begin{figure}[b!]
\hskip 3.5cm \includegraphics[width=0.45\textwidth]{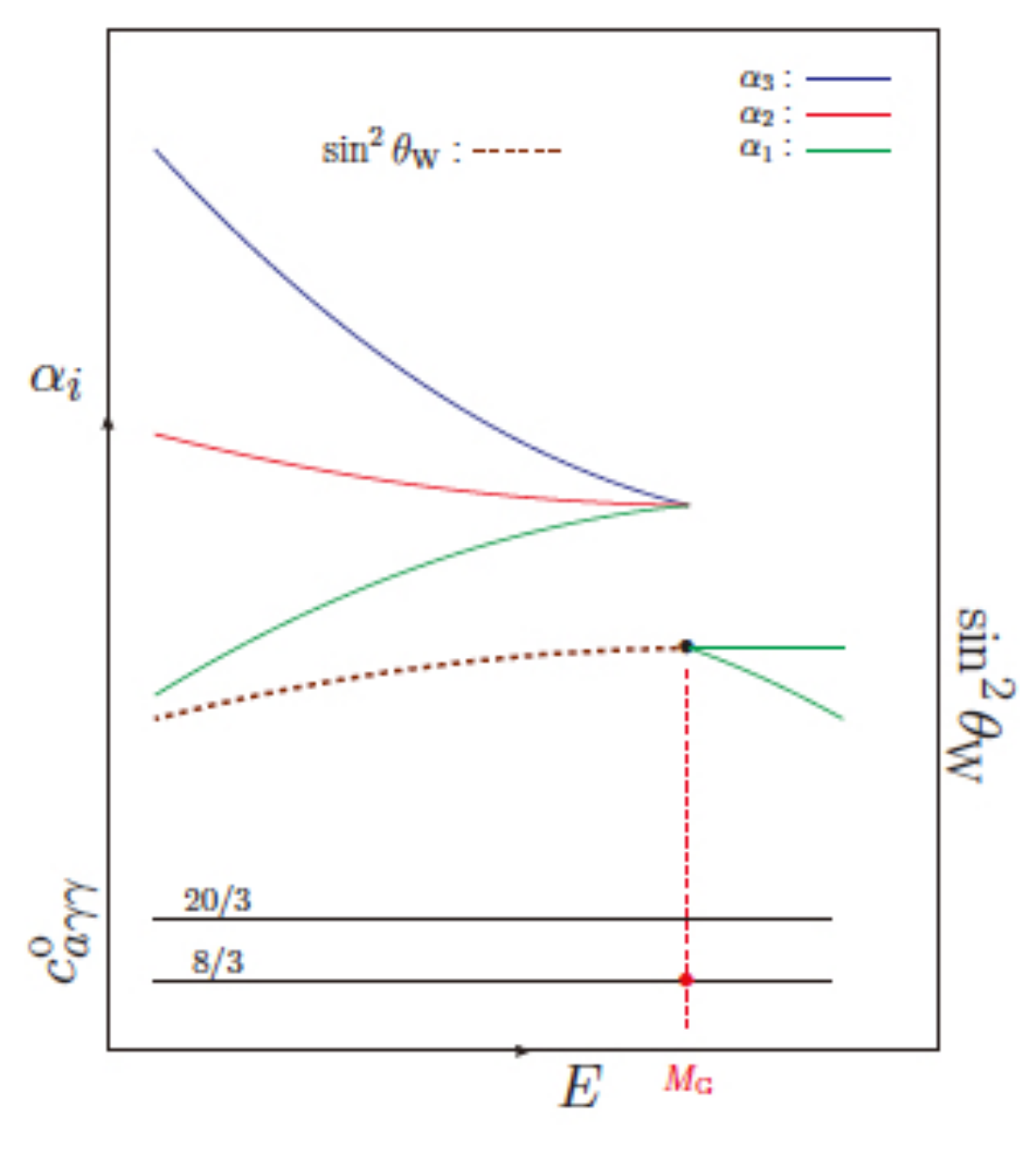}
\caption{The scale dependences of gauge couplings and $\sin^2\theta_W$.  $c_{a\gamma\gamma}^0$ does not depend on the scale.   }\label{fig:GQW}
 \end{figure}

Namely, all the other scales are negligible compared to the Planck scale. In this case, it is reasonable  to consider  that all particles which are much lighter than the Planck mass are massless in the first approximation. For fermions, ``chirality'' is the symmetry ensuring masslessness as far as it is unbroken. In the Standard Model, Weinberg used the chirality to begin with the massless fermions. In extended grand unification, Georgi formulated the criterion for GUT representations \cite{Georgi79} which is defined at the scale $M_G$ of Fig.  \ref{fig:GQW} \cite{GQW74}. Chirality is the theme of this talk. 
 In this sense, the ``invisible'' axion is closely related to GUT models. In addition, the MI axion in string theory can survive as a global symmetry down to an intermediate scale   \cite{KimPatras17}.
 
With the gauge symmetry as the only symmetry at low energy,
chiral fields and gauge bosons are the only light fields. In this regard, I attempted to concentrate my research centered around the chirality. So, there is a hope that these results derived from chirality is realized in the low energy world of Nature as in the SM \cite{Weinberg67}. So, if there exists a chiral model for fermions, then there is a good hope that it can be realized at a low energy scale. Where it is realized depends on how the chirality is broken.
In this sense, I point out a recent example based on SU(2)$\times$U(1)$_Q$ model whose factor groups are not related to any in the SM.  The fermion representations in terms of the left-handed (L-handed) chiral fields are  \cite{kim17} 
\begin{equation}
\ell_i=\left(\begin{array}{c} E_i\\ N_i \end{array}\right)_{\frac{1}{2}},E^c_{i,-1}, N^c_{i,0}, (i=1,2,3);
{\cal L}=\left(\begin{array}{c} {\cal E}\\ {\cal F} \end{array}\right)_{\frac{-3}{2}},~{\cal E}^c_{1},{\cal F}^c_{2}.
\end{equation} 

Another example is my old paper on the weak interaction singlet field $\sigma$  together with some high energy scale physics of heavy L-handed quarks $Q$ and $Q^c$ \cite{Kim79}. Probably, this was the serious one firstly going beyond the SM, proposing the {very light axion} which was later called ``invisible'' axion. This very light axion might have contributed to dark matter in the universe \cite{Preskill83}, at least some portion of it even if not the whole 27\,\% of the energy pie. The dominant part 68\,\% is dark energy which is not the issue in my talk today. Today's talk is relevan  to the remaining 5\,\%, i.e. on the abundance of atoms.

\begin{figure}[!t]
\hskip 4.5cm \includegraphics[width=0.4\textwidth]{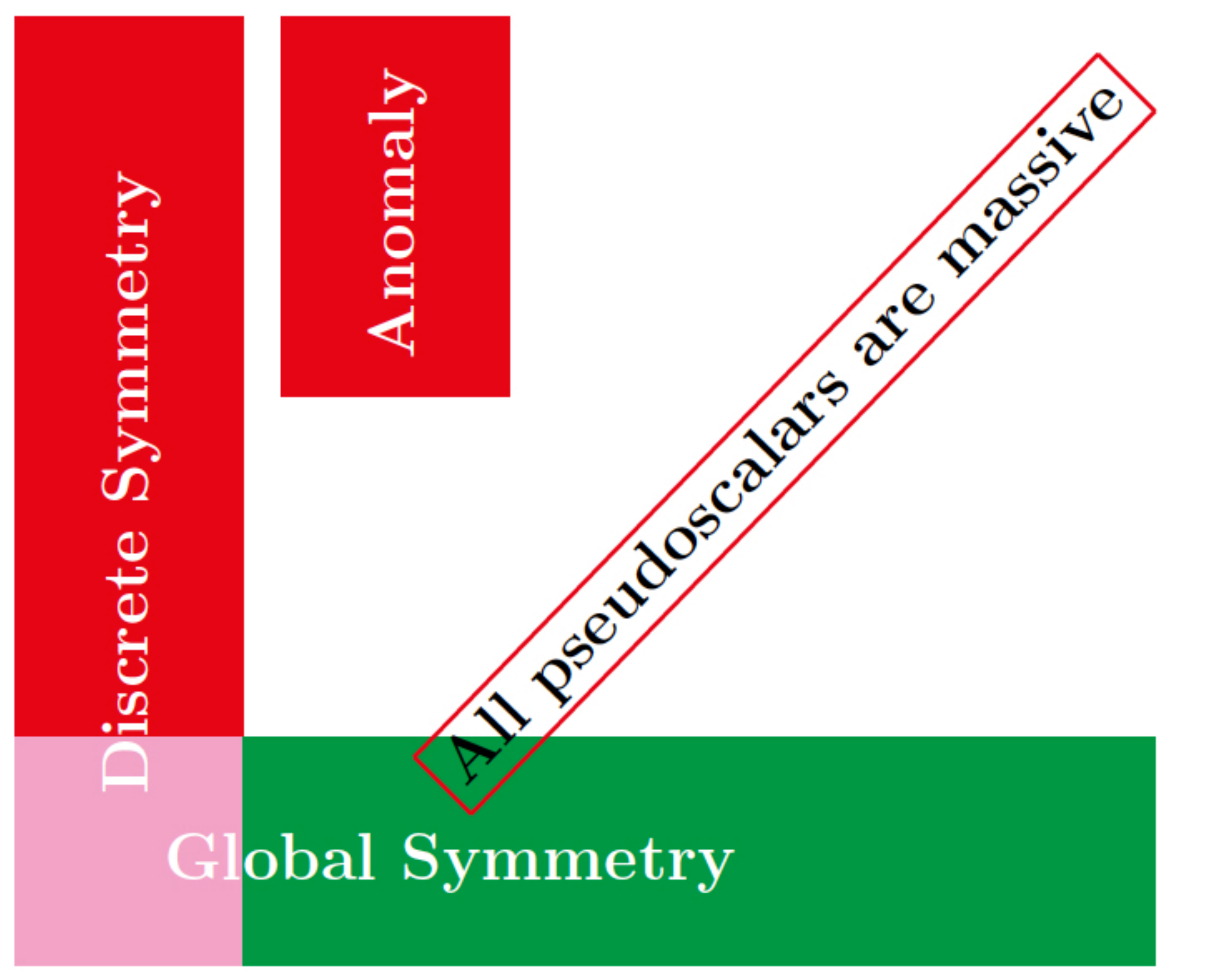} 
\caption{\label{fig:Global} A cartoon showing an accidental global symmetry from a discrete symmetry.}
\end{figure}

If we take a top-down approach such as in string compactification, global symmetries are forbidden. But some discrete symmetries can survive. In Fig. \ref{fig:Global}, this kind of discrete symmetry is symbolised  in the left lavender and red column, which includes the terms in the potential $V$. Let us consider here only a few leading terms, which is symbolised by the lavender colored terms. Then, there can result an accidental global symmetry which is symbolised by the horizontal bar including the green band.
In Fig.  \ref{fig:Global}, the red bands represent the terms breaking the global symmetry. The far left red band breaking the global symmetry is called the breaking by the terms in the potential $\Delta V$. The other red band represents the breaking of the global symmetry by  gauge anomalies. Since the global symmetry is broken anyway, all pseudoscalaers arising from breaking of the global symmetry are massive. The magnitude of the resulting mass is by the strengths of the terms in the reds. Among the anomaly contributions, the dominant one is from the QCD anomaly in the SM. If there are stronger confining force then the anomaly from that gauge group will be the dominant contribution to the mass. 
 
\begin{figure}[!t]
\includegraphics[width=0.95\textwidth]{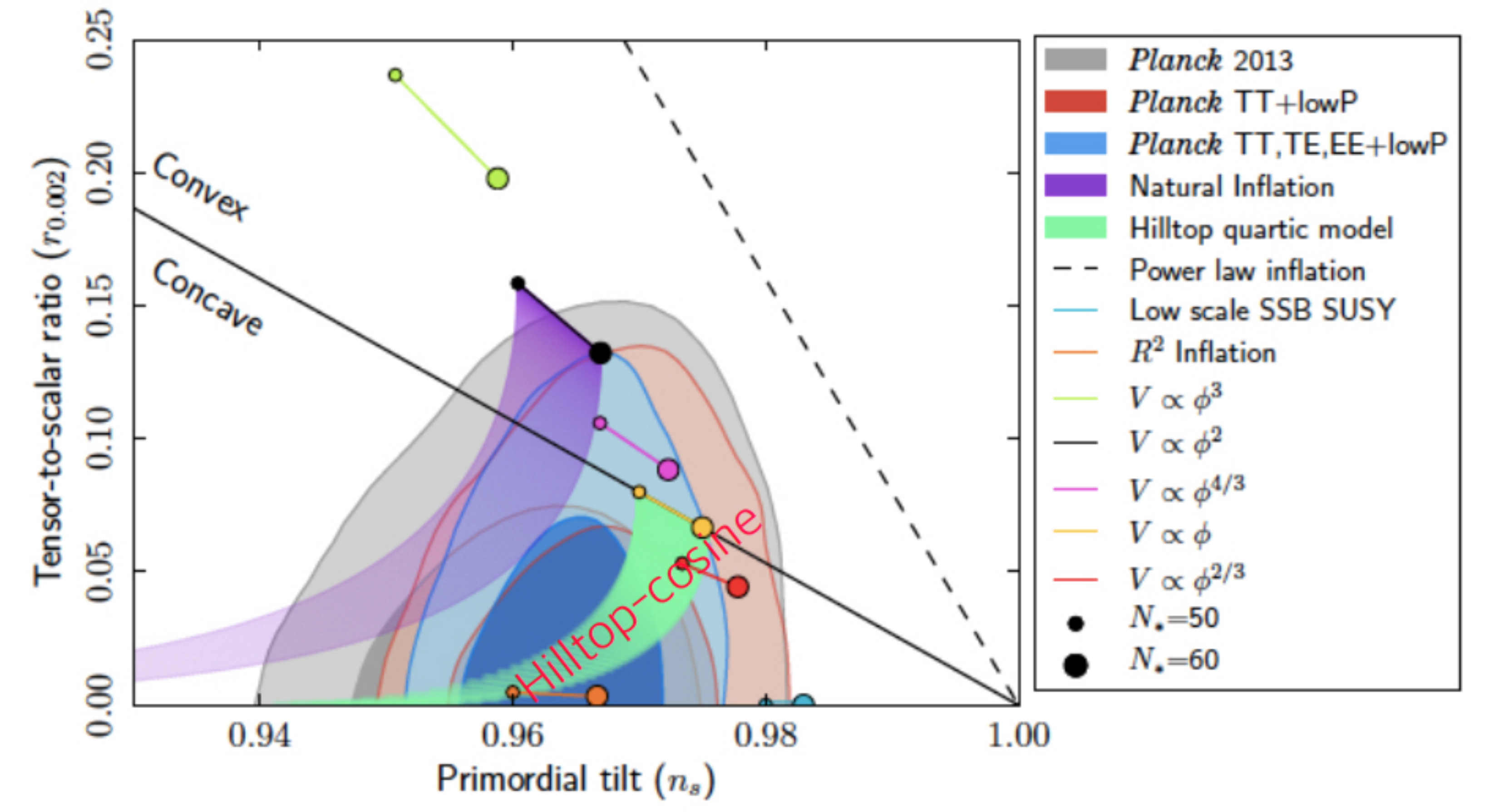} 
\caption{\label{fig:NSvsR} The hilltop-cosine inflation is in green.}
\end{figure}

Note that the hermiticity of lagrangian implies that we have potential $V=\frac12({\cal V}+{\cal V}^\dagger)$.
Thus, if we include pseudoscalars from spontaneously broken U(1), not by the anomaly term but by the potential $\Delta V$, then $180^{\rm o}$ can be a
minimum or maximum depending on the parameters. If it is a minimum, then the origin $0^{\rm o}$ can be a maximum. In this case, if we try a global symmetry for natural inflation then it is perfectly a good inflationary model. I call this {\it natural hilltop inflation} \cite{Nhilltop}. I marked this as ``Hilltop-cosine'' in the $r-n_s$ plot in Fig. \ref{fig:NSvsR}.

\section{U(1)$_{\rm anom}$: the source of ``invisible'' axion}

The mechanism behind lowering the PQ symmetry breaking scale is the so-called 't Hooft mechanism \cite{Hooft71,KimPatras17}: 
\begin{itemize}
\item[]``If a global symmetry and  a gauge symmetry are broken by the vacuum expectation value (VEV) of one complex scalar field, then the gauge symmetry is broken but a global symmetry remains unbroken''.
\end{itemize}
This is obvious because the gauge boson must obtain mass by the Brout-Englert-Higgs-Guralnik-Hagen-Kibble mechanism and one continuous shift symmetry from the original two angle directions cannot be broken. This unbroken shift symmetry is a global symmetry because there does not exist the  corresponding gauge boson below the VEV. This fact was noted long time ago
 \cite{Kim88} in string compactification.

  The signal for the existence of gauge boson mass in spontaneously broken gauge models arises from the mixing term of the longitudinal mode $a$ (the phase of the complex scalar field $\phi$ used in the spontaneous symmetry breaking) and the gauge field $A_\mu$
 \begin{equation}
 |D_\mu \phi|^2=\frac{1}{2}(\partial_\mu a)^2-g Q_a A_\mu\partial^\mu a+\frac{g^2}{2}Q_a^2 v^2 A_\mu^2
 =\frac{g^2}{2}Q_a^2 v^2(A_\mu- \frac{1}{gQ_a v}\partial_\mu a)^2 
\end{equation}
and the longitudinal degree disappears by redefining the longitudinal component of $A_\mu$ as  $A_\mu'=A_\mu-  ({1}/{gQ_a v})\partial_\mu a$.
 
In the compactification of the heterotic string, 10D$\to$4D, the model-independent (MI) axion component comes from the tangential component of $B_{MN}(M,N=1,2,\cdots,10)$  \cite{GS84,Witten84}: $B_{\mu \nu}(\mu,\nu=1,2,3,4)$,
 \begin{equation}
 H_{\mu\nu\rho}= M_{MI}\epsilon_{\mu\nu\rho\sigma}\partial^\sigma a_{MI}.
\end{equation}
The rank 16 gauge group E$_8\times$E$_8'$ can produce many U(1)'s beyond the SM gauge group. If all U(1) do not have any gauge anomaly, then the shift symmetry $a_{MI}\to a_{MI}+$(constant) is broken at the compactification scale. On the other hand, if there appears an anomalous U(1) from E$_8\times$E$_8'$ \cite{AnomU187},  U(1)$_{\rm anom}$, then the gauge boson $A_\mu^{\rm anom}$ corresponding to U(1)$_{\rm anom}$ obtains mass.  Here,  $H_{\mu\nu\rho}$ couples to the anomalous gauge boson $M_{MI}A_\mu^{\rm anom} \partial^\mu a_{MI}$ and the 't Hooft mechanism works as shown in   \cite{KimKyaeNam17}. So, in the string compactification models with a 4D anomalous U(1) gauge symmetry, the original shift symmetry of the MI axion, $a_{MI}\to a_{MI}+$(constant), survives as a global PQ symmetry below the compactification scale. The ``invisible'' axion realized around $10^{10\sim 11\,}$GeV scale can have this origin of the PQ symmetry from string compactification.

Let me comment on the allowed magnitude on the $\theta_{\rm QCD}$ parameter in QCD.  Since we missed a factor $g_{\pi NN}$ in our ten years old review on axions,  an erratum just for this missed factor  was written recently \cite{KimRMP19}. The correct value is
\begin{eqnarray}
\frac{d_n}{e}=\frac{g_{\pi nn}\overline{g_{\pi nn}}}{4\pi^2 m_n}\ln \left(\frac{m_n}{m_\pi} \right)=\frac{2g^2_{\pi nn} }{4\pi^2 m_n}\frac{|\theta_{\rm QCD}|}{3}\ln \left(\frac{m_n}{m_\pi} \right) \nonumber\\[0.3em]
=\frac{3.60}{m_n}|\theta_{\rm QCD}|
\end{eqnarray}
which leads to
\begin{equation}
|\theta_{\rm QCD}|\lesssim 2.8\times 10^{-13}.
\end{equation}
This is calculated with only one family of quarks and we did not introduce the strange quark, unlike in Ref. \cite{Crewther79}. This small value of $\theta_{\rm QCD}$ has a difficulty in the Nelson--Barr type calculable solutions \cite{BarrNelson}.

Before discussing the gauge hierarchy, let me point out the intermediate scales in the KSVZ and in the DFSZ models. In the KSVZ model, a renormalizable lagrangian with a heavy quark $Q$ is introduced,
\begin{equation}
\textrm{KSVZ:}~{\cal L}_Q=-f\overline{Q}_LQ_R\sigma+{\rm h.c}.
\end{equation}
which violates the Peccei--Quinn symmetry by the QCD anomaly term. Here $\langle \sigma\rangle $ is at the intermediate scale $10^{10}-10^{12\,}$GeV. On the other hand, the DFSZ model may introduce a following renormalizable lagrangian
\begin{equation}
\textrm{DFSZ:}~V=\frac{\lambda}{4} (\sigma^*\sigma)^2-\frac{\mu^2}{2}(\sigma^*\sigma)+\lambda_1\sigma^2 H_u H_d +{\rm h.c}.\label{eq:DFSZV}
\end{equation}
where $H_u$ and $H_d$ are Higgs doublets giving mass to $Q_{\rm em}=\frac{+2}{3}, \frac{-1}{3}$ quarks, respectively. Because the VEVs of  $H_u$ and $H_d$ of order the electroweak scale and the VEV of $\sigma$ is of order the intermediate scale, $\frac{\lambda}{\lambda_1}$ must be of order again $10^{-9}$. This is a hierarchy.  An axion was  introduced to solve the smallness problem of  $\theta_{\rm QCD}$, but another hierarchy of couplings needed in Eq. (\ref{eq:DFSZV})  is of that order again. Introduction of supersymmetry (SUSY), however, avoids this problem by a non-renormalizable $\mu$ term \cite{KimNilles84},
\begin{equation}
\frac{H_uH_d}{M}  \sigma\sigma + {\rm h.c}.
\end{equation}
\begin{figure}
\hskip 2.7cm\includegraphics[width=0.65\textwidth]{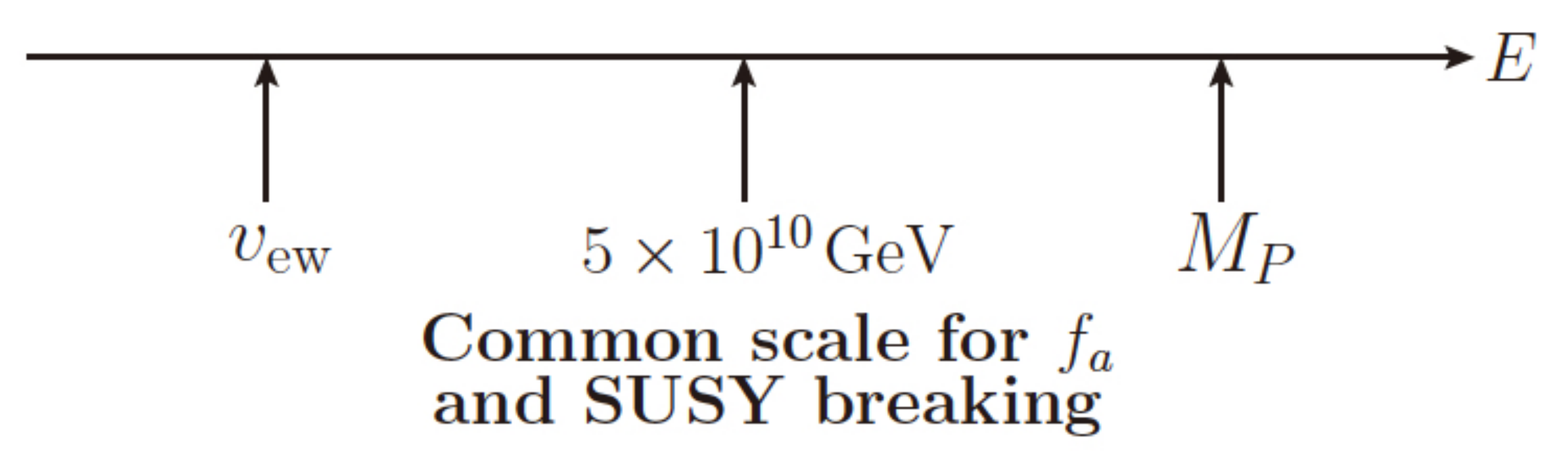} 
\caption{\label{fig:Common} The common intermediate scale.}
\end{figure}
Now, the question is how  we determine the VEV of the singlet field $\sigma$. It is related to a solution of the gauge hierarchy problem. We had an ansatz for the common scale \cite{Common} for  the VEV of  $\sigma$, where the scale of SUSY breaking is near the point of the squareroot of the Planck mass and the electroweak scale $v_{\rm ew}$, as depicted in Fig. \ref{fig:Common}.

\section{\label{sec:GaugeH}Gauge hierarchy}

With respect to the GUT and Planck scales, the electroweak scale has the hierarchy of $10^{-14}$ and $10^{-16}$, respectively. After the advent of GUTs, the need for spontaneous breaking of the SU(5) GUT, for example, is given by the potential
\begin{equation}
V=-M^2 \Sigma^\dagger\Sigma -v_{\rm ew}^2H^\dagger H+\cdots
\end{equation}
where $\Sigma$ is {\bf 24} needed for breaking SU(5) and $H$ is {\bf 5} containing the Higgs doublet of the SM.  The needed parameters $v_{\rm ew}^2$ and $M^2$ in the potential must be tuned to a ratio of order $10^{-28}$, which constitutes the essence of the gauge hierarchy problem. Why is there such an extreme ratio of parameters? These are on the scalar masses and hence the cutoff is usually taken around TeV.
  
\begin{figure}
\hskip 4cm\includegraphics[width=0.45\textwidth]{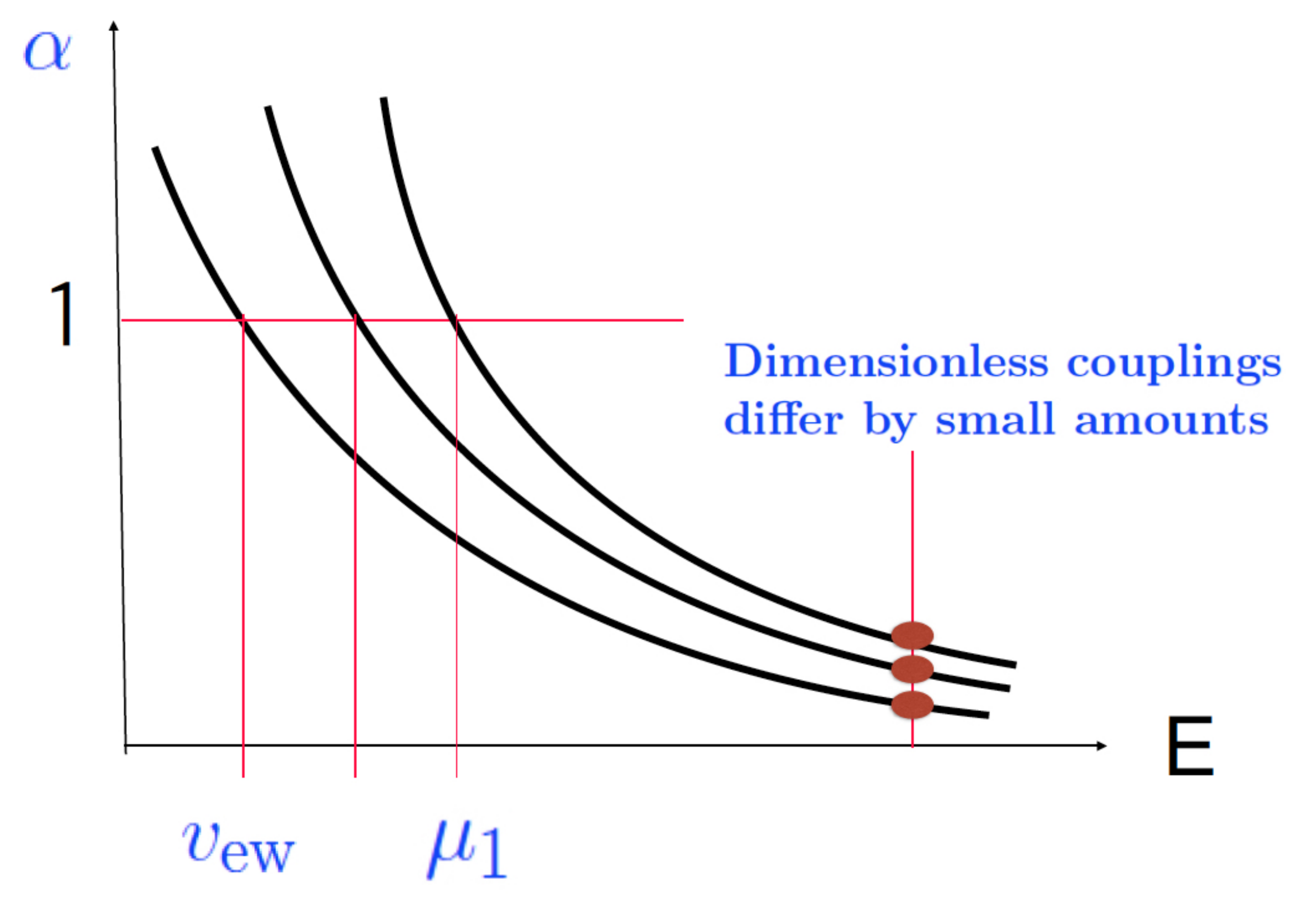} 
\caption{\label{fig:DimTrans} The idea of dimensional transmutation.}
\end{figure}

An exponential hierarchy is desirable, which can be obtained  by dimensional transmutation with a confining (asymptotically free) nonabelian gauge group. This idea of dimensional transmutation is depicted in Fig. \ref{fig:DimTrans}. The values at the brown marks give dimensionless numbers on the coupling constants. Even if the coupling constants differ by small amounts, the asymptotic freedom gives some difference in coupling counstants at exponentially different scales, such as  $v_{\rm ew}$ and $\mu_1$ in the figure, that are defined to be the scale where the coupling constant are of order 1.

This idea was used in late 1970's under the name of technicolor. For the technicolor to distinguish families, the flavor group must be included. In this kind of cmxtended technicolor theories,  the precision data  are not consistent with the idea and the technicolor was   avoided since then. 

To give masses to the SM fermions, Higgs scalars are needed.  Also, the LHC data hint that the Higgs boson couplings are proportional to the fermion masses. For the flavor, therefore, Higgs scalars are definitely needed. But, the VEV scale at $v_{\rm ew}$ introduces the aforementioned hierarchy problem. For a small VEV of $H$, firstly one has to introduce $H$ as a massless scalar. Then, allow it to develop a VEV at the electroweak scale.
 Fermions can be light if an appropriate chiral property is given. For scalars, there is no such chiral symmetry because scalars do not have a non-vanishing spin. The reason is the following.
 
For a composite nucleus-electron with a non-zero orbital angular momentum, one can calculate the orbital angular momentum as ${\bf L}={\bf r}\times {\bf p}$. Let us choose the $+\hat{z}$ as the propagating direction. If ${\bf r}=0$ then the orbital angular momentum is zero. If the composite has a nonzero angular momentum, then the particle moving into the  $+\hat{z}$ direction has the orbiting plane perpendicular to $\hat{z}$.
Generalizing this, any non-zero spin particle  has two transverse degrees if it moves with light velocity. If it has mass, the velocity can be smaller than the light velocity,   and this argument does not apply. If a particles moves in the $\hat{z}$ direction only, these two transverse degrees for a massless particle do not apply.  Composite scalars are not carrying the orbital angular momentum, and it must move in the $\hat{z}$ axis. Generalizing this,  for scalars moving in the  $\hat{z}$ axis, there is no reason that it should move with the light velocity. Therefore, the natural mass scale of a scalar is the scale where the scalar is defined.

Some extra symmetry is needed to makes the scalar  light. For this reason, SUSY has been used for Higgs doublets to carry a kind of chirality through their superpartner fermions.
 To assign a chirality to a scalar  or for the absence of quadratic divergene, the N=1 SUSY was introduced in particle physics phenomenology. In the last three decades, the technical problem  on the relative parameter scales in the SUSY models were emphasized, for example,
 \begin{eqnarray}
 \delta m_h^2 =\frac{3G_F^2}{4\sqrt2\,\pi^2} \left(4m_t^2-2m_W^2-m_Z^2-m_h^2 \right)\Lambda^2,\label{eq:delmh}\\
 \frac{M_Z^2}{2}=\frac{m^2_{H_d}-m^2_{H_u}\tan^2\beta}{\tan^2\beta-1}-\mu^2.\label{eq:MZ}
 \end{eqnarray}
 where $\Lambda$ is the cutoff scale. Since top quark is much heavier than other particles, which is the dominent contribution in Eq. (\ref{eq:delmh}), and $\Lambda$ is basically the cutoff which is identified as the superpartner mass scale. For a dimensionless $\delta m_h^2$ of O(1),
 $(0.94\times 10^{-3}{\rm GeV}^{-1}\,\Lambda)^2$ is of order 1, and the cutoff scale is of order TeV. This is the phenomenological scenario for the superpartner masses. In terms of the parameters, e.g. the masses of two Higgs doublets and the $\mu$ parameter which are of order $\Lambda$, the $Z$ boson mass square of  0.01 TeV$^2$ should result. If the superpartner masses are large, this introduces another fine-tuning called the {\it little hierarchy}. A little hierarchy of 1\,\% is generally accepted.
 
Some here might have worked on standard-like models from superstring, not worrying about the gauge symmetry breaking  at the GUT scale. The reason that the standard-like models are attractive is that they are chiral models and that there is no need for  further breaking of gauge symmetries down to the SM. The SUSY breaking in supergravity needs an intermediate scale for SUSY breaking as depicted in Fig. \ref{fig:Common}. As glimpsed in Fig. \ref{fig:DimTrans}, we need SUSY breaking by dimensional transmutation.  To break SUSY dynamically was known to be very difficult \cite{Witten81}, and hence gravity intervention from superstring got a lot of interest \cite{DIN85}.  Here we do not borrow the interference from gravity or from string theory. We will look for dynamical SUSY breaking just from a confining force.
 
As Figs.  \ref{fig:Common} and \ref{fig:DimTrans} show, the intermediate scale is quite small compared to the Planck mass. Therefore, a chiral model is needed to bring down the spectrum to a low energy scale, i.e. to an intermeiate scale. But for consistency, anomaly freedom is required in the effective theory. Howard Georgi formulated \cite{Georgi79} low energy effective theory with the hypothesis: {\it SURVIVAL HYPOTHESIS}. The survival hypothesis in gauge theories requires only the chirality.  If it is decorated such as by discrete or global symmetries, then there are practically uncountable possibilities and hence it does not have any predictive power. Georgi also extended `fundamental representation' to include all antisymmetric representations such that quarks and anti-quraks are only color {\bf 3} and $ \overline{\bf 3}$. In this scheme, SU(3) does not allow any chiral representation. Also, SU(4)  does not allow any chiral representation. The smallest gauge group allowing a chiral representation is SU(5) Georgi--Glashow model \cite{GG74}, with the representations,
\begin{equation}
{\bf 10}\equiv [2],~~{\overline{\bf 5}}\equiv [4],\label{eq:GGmodel}
\end{equation}
which does not have SU(5) gauge anomaly.
 
\section{\label{sec:MI} Generation of $M_I$ by dimensional transmutation}

In the last two decades, SUSY QCD, {\it i.e.} gauge theories with vector-like representations, was extensively studied  mainly to understand `duality' concept. Because they are vector-like, they are not useful for our chiral case. Now we look for chiral models for SUSY breaking.

\begin{figure}
\hskip 3.8cm\includegraphics[width=0.45\textwidth]{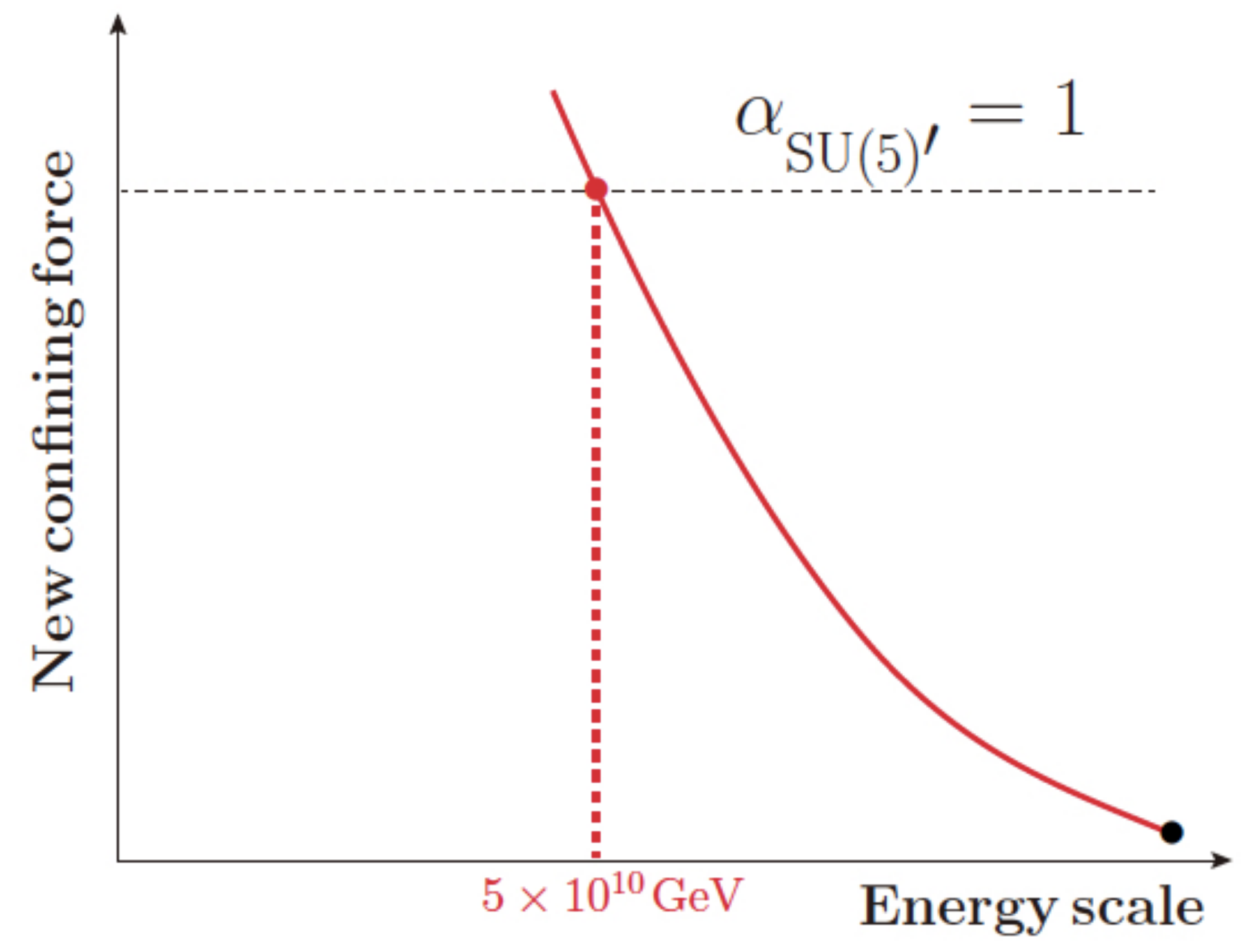} 
\caption{\label{fig:SU5} Evolution of the gauge coupling constant of  hidden sector SU(5)$'$.}
\end{figure}

Namely, we search for chiral models in SUSY GUT (SGUT) models. Indeed, this was performed by Meurice and Veneziano with the one-family Georgi--Glashow model \cite{Meurice84}. We anticipate that the hidden sector SU(5)$'$ confines at an intermediate scale as shown in Fig. \ref{fig:SU5}. But Meurice and Veneziano could not pursue any further because they could not write a superpotential with the terms in Eq. (\ref{eq:GGmodel}), even though they predicted ``In the future further calculations should not fail to provide a complete systematics of the circumstances under which spontaneous SUSY breaking takes place.''

It took 35 years to close this loop in our paper \cite{KimKyae19}. The hidden sector SU(5)$'$ representations under  the group (SU(5)$_{\rm gauge}'$, SU(2)$_{\rm global}$) are
 \begin{eqnarray}
\Psi^{\alpha\beta} \oplus ~\overline{\psi}^\alpha_1~\oplus 2\cdot \psi_{2\,\alpha}\nonumber \\
(\overline{\bf 10},\bf 1) \oplus({\overline{\bf 5}}, \bf 1)\oplus(\bf 5,2),\label{eq:KKmodel}
\end{eqnarray}
which does not have an SU(5)$_{\rm gauge}'$ anomaly. We were guided to this representation from a compactification of heterotic string  of \cite{KimKyae09}. Now we can write the following superpotential terms
\begin{equation}
W_0\ni \frac14 \overline{\Psi}^{\,\alpha\beta}\psi^i_{2\alpha}\psi^j_{2\beta}\varepsilon_{ij},~\overline{\psi}^{\,\alpha}_1\psi^i_{2\alpha} D_{1i},~\frac{1}{5!} \overline{\Psi}^{\,\alpha\beta} \overline{\Psi}^{\,\gamma\delta} \overline{\psi}^{\,\epsilon}_1 \varepsilon_{\alpha\beta\gamma\delta\epsilon}.
\end{equation}
These three couplings work as conditions and there remain one global symmetry U(1)$_{\rm global}$. Below the SU(5)$'$ confinement scale, we can consider the following  SU(5)$'$ singlets,
\begin{equation}
\phi=\frac{1}{5!} \overline{\Psi}^{\,\alpha\beta} \overline{\Psi}^{\,\gamma\delta} \overline{\psi}^{\,\epsilon}_1 \varepsilon_{\alpha\beta\gamma\delta\epsilon},~~\Phi_i=\overline{\psi}^{\,\alpha}_1\psi_{2\alpha\,i}.
\end{equation}

Let us consider the anomaly U(1)$_{\rm global}$--SU(2)$_{\rm gauge}$--SU(2)$_{\rm gauge}$ around the confinement scale. As in axion physics, the $\theta$ term anomaly appears, due to instanton effects. If we consider a very large instanton, effectively an infinite size, the anomaly can be  U(1)$_{\rm global}$--SU(2)$_{\rm global}$--SU(2)$_{\rm global}$. This is an interpretation of the global anomaly matching condition of 't Hooft \cite{Hooft79}. For  U(1)$_{\rm global}$--U(1)$_{\rm gauge}$--U(1)$_{\rm gauge}$, we do not have the instanton argument and we need not satisfy the matching of U(1)$_{\rm global}$--U(1)$_{\rm global}$--U(1)$_{\rm global}$. Even if the $\theta$ term is written with the constant angle $\theta$,  $\sim \frac{\theta}{32\pi^2}F_{\mu\nu} \tilde{F}^{\mu\nu}$, it is a total derivative,
\begin{eqnarray}
&&\theta \frac14 \varepsilon^{\mu\nu\rho\sigma} (\partial_\mu A_\nu-\partial_\nu A_\mu)(\partial_\rho A_\sigma-\partial_\sigma A_\rho)\nonumber\\[0.3em]
&& =\theta  \varepsilon^{\mu\nu\rho\sigma} (\partial_\mu A_\nu)(\partial_\rho A_\sigma )\nonumber
\\[0.3em]
&&=\partial_\rho [\theta  \varepsilon^{\mu\nu\rho\sigma}(\partial_\mu A_\nu)(A_\sigma)] -[\partial_\rho (\theta  \varepsilon^{\mu\nu\rho\sigma}\partial_\mu A_\nu)](A_\sigma)\nonumber\\[0.3em]
&&=\partial_\rho [\theta  \varepsilon^{\mu\nu\rho\sigma}(\partial_\mu A_\nu)(A_\sigma)] -[(\theta  \varepsilon^{\mu\nu\rho\sigma}\partial_\rho \partial_\mu A_\nu)](A_\sigma))\nonumber\\[0.3em]
&&=\partial_\rho [\theta  \varepsilon^{\mu\nu\rho\sigma}(\partial_\mu A_\nu)(A_\sigma)] ,
\end{eqnarray}
and hence can be neglected in the action.
In Table \ref{tab:Matching}, we list the quantum 
\begin{table}
\caption{\label{tab:Matching}This table is Table 1 of Ref. \cite{KimKyae19}.}
\begin{tabular}{c|c|ccc|cc}
 &&&&&&\\[-1.3em]  
 & SU(2)&U(1)$_{\overline{\Psi}}$&U(1)$_{\overline{\psi}_1}$&U(1)$_{{\psi}_2}$ &U(1)$_{AF}$&U(1)$_{R}$\\ \hline
   $\vartheta$ &$0$ &$0$&$0$  &$0$&$0$&$+1$   \\[0.3em]
 $\overline{\Psi}\sim(\bf \overline{10}, 1)$ 
 &${\bf 1}$&$+1$&$0$&$0$& $-1$&$+\frac12$ \\[0.3em]
 fermion &${\bf 1}$  &$+1$ &$0$ &$0$ &$-1$&$-\frac12$ \\[0.3em]
$\overline{\psi}_1\sim(\bf \overline{5}_1, 1)$ 
 &${\bf 1}$&$0$&$+1$&$0$& $+2$&$+1$ \\[0.3em]
 fermion &${\bf 1}$  &$0$ &$+1$ &$0$ & $+2$&$0$ \\[0.3em]
${\psi}_2\sim(\bf {5}, 2)$ 
 &${\bf 2}$&$0$&$0$&$+1$& $+\frac12$&$+1$ \\[0.3em]
 fermion &${\bf 2}$  &$0$ &$0$ &$+1$  &$+\frac12$&$0$ \\[0.3em]
$D\sim(\bf {1}, 2)$ 
 &${\bf 2}$&$0$&$0$&$0$& $-\frac52$&$0$ \\[0.3em]
 fermion &${\bf 2}$  &$0$ &$0$ &$0$   &$-\frac52$&$-1$ \\[0.3em]
$W^a\sim\lambda^a$ 
 &$--$&$0$&$0$&$0$&$0$ &$+1$ \\[0.3em]
 $\Lambda^b$& $--$  &$--$  &$--$ &$--$ &$--$&$\frac{2b}{3}$ \\[0.3em]
 \hline
${\phi}$ 
&${\bf 1}$&$--$&$--$&$--$& $-5$&$+2$ \\[0.3em]
fermion &${\bf 1}$  &$--$ &$--$ &$--$ & $-5$&$+1$ \\[0.3em]
${\Phi}_i$ 
&${\bf 2}$&$--$&$--$&$--$& $+\frac52$&$+2$ \\[0.3em]
fermion &${\bf 2}$  &$--$ &$--$ &$--$ & $+\frac52$&$+1$ \\[0.3em]
$S$ 
&${\bf 1}$&$0$&$0$&$0$& $0$&$+2$ \\[0.3em]
fermion &${\bf 1}$  &$0$ & $0$ &$0$ &$0$&$+1$ \\[0.3em]
$D_i\sim(\bf {1}, 2)$ 
 &${\bf 2}$&$--$&$--$&$--$& $-\frac52$&$0$ \\[0.3em]
 fermion &${\bf 2}$  &$--$ &$--$ &$--$   &$-\frac52$&$-1$ \\[0.3em]
\end{tabular}
\end{table}
numbers above and below the confinement scale. For global U(1)'s, we also listed the U(1)$_R$ charges of SUSY theory. The global U(1) we mentioned before is U(1)$_{AF}$ which is anomaly free above the confinement scale. So, there is no problem with U(1)$_{AF}$. Because of the U(1)$_R$ symmetry which gives two units to $W$, we need not consider other composite particles beyond those listed in Table  \ref{tab:Matching}, $\phi, \Phi_i$ and the gluino condensation $S$. The extra higher dimensional composites must have $R>2$.

We have the following superpotential in terms of composite fields,
\begin{equation}
W= M^2\phi +\frac{N_c(N_c^2-1)}{32\pi^2}\mu_0^2 S\left(1-a\log\frac{\Lambda^3}{S\mu_0^2} \right) +b M\Phi_i D^i.\label{eq:CompW}
\end{equation}
Below the confinement scale, there is no other terms in the superpotential. From Eq. (\ref{eq:CompW}),
we have the following SUSY conditions,
\begin{eqnarray}
&&\frac{\partial W}{\partial\phi}=0
\to M^2=0,\\[0.3em]
&&\frac{\partial W}{\partial\Phi_i}=0
\to D^i=0,\\[0.3em]
&&\frac{\partial W}{\partial\phi}=0
\to \Phi_i=0,\\[0.3em]
&&\frac{\partial W}{\partial\phi}=0
\to \mu_0^2 \left(1+a-a\log\frac{\Lambda^3}{S\mu_0^2} \right) =0.
\end{eqnarray}
 If we define the coupling $\lambda_0$ above the confinement scale as,\begin{equation}
 \frac{\lambda^0}{5!} \overline{\Psi}^{\,\alpha\beta} \overline{\Psi}^{\,\gamma\delta}\overline{\psi}^{\,\epsilon}_1 \varepsilon_{\alpha\beta\gamma\delta\epsilon}\to \lambda_0\mu_0^2\phi=M^2\phi,
\end{equation}
then  $M^2$ is nonzero since $\lambda_0$ is defined to be nonzero. Here $\mu_0$ is a scale introduced at the confinement point $\Lambda$. Therefore,  SUSY is broken by the 'O Raifeartaigh mechanism.

\begin{figure}
\hskip 3cm\includegraphics[width=0.65\textwidth]{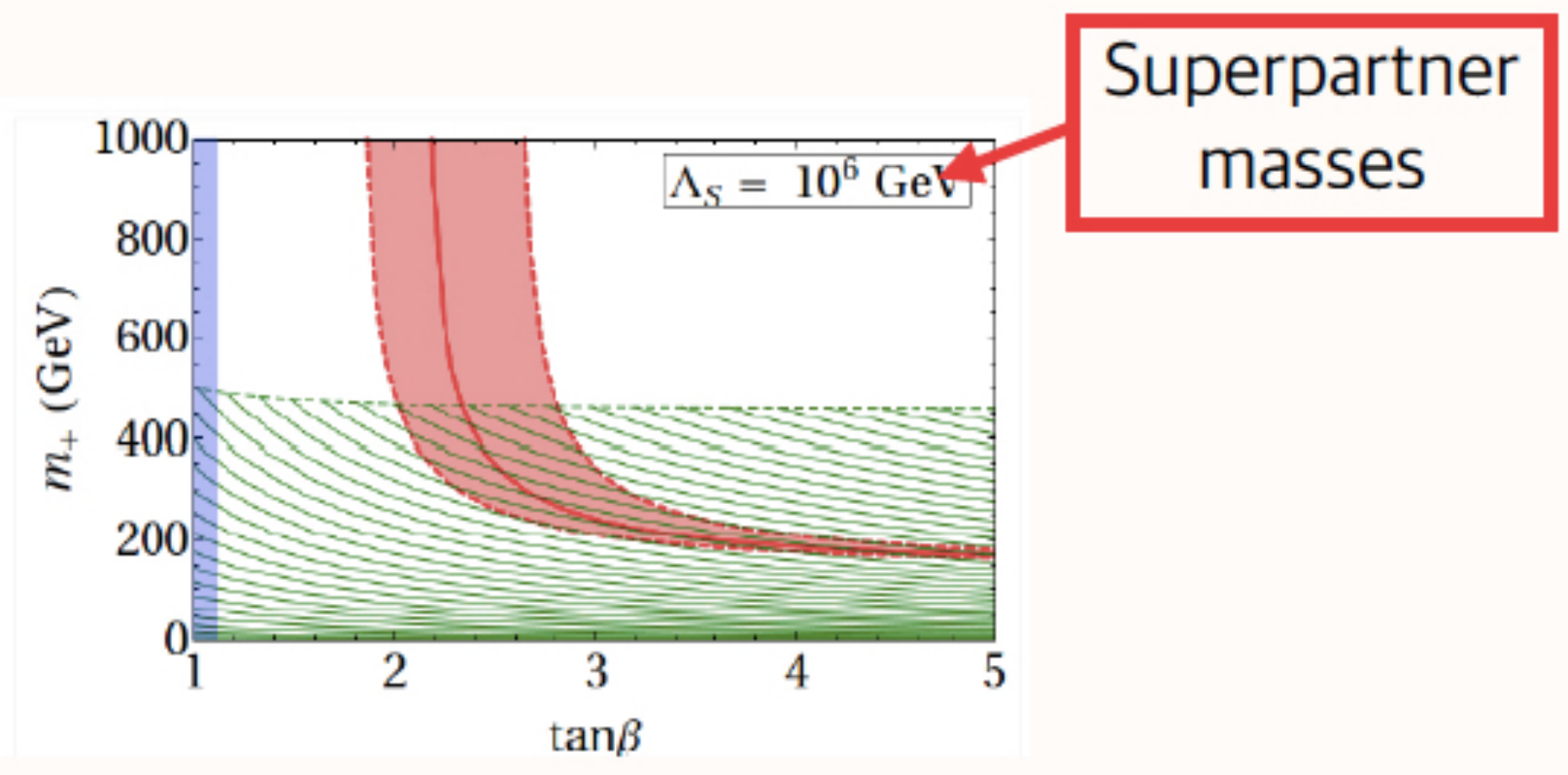} 
\caption{\label{fig:Corfu} Allowed region on the cutoff scale.}
\end{figure}

If the hidden SU(5)' confines at $5\times 10^{10}$ GeV -- $10^{12}$
GeV, the SUSY breaking
scale for the SM partners is above 1 TeV.
In particular, the lower end  $5\times 10^{10}$
--  $10^{11}$
GeV is particularly interesting
because it is the anticipated axion scale envisioned in Fig. \ref{fig:Common}, which however is experimentally the most difficult region for the axion search.
The SU(5)$'$ confinement provides this region because of the composite-scalar ($\phi$) condensation, rather than gaugino condensation.

In our case, the confinement scale by the singlet
composite scalar is somewhere between  $5\times 10^{10}$
GeV -- $10^{12}$ GeV, but not as high as  $10^{13}$ GeV.
With this, $M_{\rm SUSY}$ can be raised to the scale of the so-called little hierarchy. For a free parameter  $a$  larger than 1, the superpartner scale at $a\,$TeV
needs   $\sqrt{a}\cdot 5\cdot 10^{10}$ GeV for the confinement scale.
 For a free parameter  $a$  larger than 1, the superpartner scale at $a\,$TeV
needs   $\sqrt{a}\cdot 5\cdot 10^{10}$ GeV for the confinement scale. A 6 TeV superpartner mass  needs  the confinement scale at $10^{11}$ GeV.
Indeed, this can be working as a talk \cite{Saha19} at the Corfu Workshop showed Fig. \ref{fig:Corfu} where the superpartner masses around 6 TeV and charged Higgs scalar masses at several hundred GeV are allowed. In that study, of course, the Higgs boson mass  at 125 GeV was used as a condition.

\begin{table}
\caption{\label{tab:Matching}This table is Table 1 of Ref. \cite{KimKyae19}.}
\begin{tabular}{c|c|ccc|cc}
 &&&&&&\\[-1.3em]  
 & SU(2)&U(1)$_{\overline{\Psi}}$&U(1)$_{\overline{\psi}_1}$&U(1)$_{{\psi}_2}$ &U(1)$_{AF}$&U(1)$_{R}$\\ \hline
   $\vartheta$ &$0$ &$0$&$0$  &$0$&$0$&$+1$   \\[0.3em]
 $\overline{\Psi}\sim(\bf \overline{10}, 1)$ 
 &${\bf 1}$&$+1$&$0$&$0$& $-1$&$+\frac12$ \\[0.3em]
 fermion &${\bf 1}$  &$+1$ &$0$ &$0$ &$-1$&$-\frac12$ \\[0.3em]
$\overline{\psi}_1\sim(\bf \overline{5}_1, 1)$ 
 &${\bf 1}$&$0$&$+1$&$0$& $+2$&$+1$ \\[0.3em]
 fermion &${\bf 1}$  &$0$ &$+1$ &$0$ & $+2$&$0$ \\[0.3em]
${\psi}_2\sim(\bf {5}, 2)$ 
 &${\bf 2}$&$0$&$0$&$+1$& $+\frac12$&$+1$ \\[0.3em]
 fermion &${\bf 2}$  &$0$ &$0$ &$+1$  &$+\frac12$&$0$ \\[0.3em]
$D\sim(\bf {1}, 2)$ 
 &${\bf 2}$&$0$&$0$&$0$& $-\frac52$&$0$ \\[0.3em]
 fermion &${\bf 2}$  &$0$ &$0$ &$0$   &$-\frac52$&$-1$ \\[0.3em]
$W^a\sim\lambda^a$ 
 &$--$&$0$&$0$&$0$&$0$ &$+1$ \\[0.3em]
 $\Lambda^b$& $--$  &$--$  &$--$ &$--$ &$--$&$\frac{2b}{3}$ \\[0.3em]
 \hline
${\phi}$ 
&${\bf 1}$&$--$&$--$&$--$& $-5$&$+2$ \\[0.3em]
fermion &${\bf 1}$  &$--$ &$--$ &$--$ & $-5$&$+1$ \\[0.3em]
${\Phi}_i$ 
&${\bf 2}$&$--$&$--$&$--$& $+\frac52$&$+2$ \\[0.3em]
fermion &${\bf 2}$  &$--$ &$--$ &$--$ & $+\frac52$&$+1$ \\[0.3em]
$S$ 
&${\bf 1}$&$0$&$0$&$0$& $0$&$+2$ \\[0.3em]
fermion &${\bf 1}$  &$0$ & $0$ &$0$ &$0$&$+1$ \\[0.3em]
$D_i\sim(\bf {1}, 2)$ 
 &${\bf 2}$&$--$&$--$&$--$& $-\frac52$&$0$ \\[0.3em]
 fermion &${\bf 2}$  &$--$ &$--$ &$--$   &$-\frac52$&$-1$ \\[0.3em]
\end{tabular}
\end{table}
In Table \ref{tab:Matching}, we list the quantum 
numbers above and below the confinement scale. For global U(1)'s, we also listed the U(1)$_R$ charges of SUSY theory. The global U(1) we mentioned before is U(1)$_{AF}$ which is anomaly free above the confinement scale. So, there is no problem with U(1)$_{AF}$. Because of the U(1)$_R$ symmetry which gives two units to $W$, we need not consider other composite particles beyond those listed in Table  \ref{tab:Matching}, $\phi, \Phi_i$ and the gluino condensation $S$. The extra higher dimensional composites must have $R>2$, and hence are forbidden from SUSY.
 
\section{Conclusion}

In conclusion, I talked {\it chirality} for low mass particles and dynamical SUSY breaking. It can
  solve the difficult problem of gauge
hierarchy by an SU(5)$'$ confining group with a specific representation. Actually, such a spectrum can arise from string compactification.


\begin{thebibliography}{99}
\bibitem{Nhilltop} Anyway, in all citations of natural inflations, I do not find the corresponding nonabelian gauge groups in the papers and talks.
\end{thebibliography}
\end{document}